\documentclass[10pt,letterpaper]{article}
\usepackage{opex3}

\usepackage{bm}

\usepackage{amsmath}
\usepackage{amssymb}

\usepackage[utf8]{inputenc}

\usepackage{units}
\usepackage[amssymb, cdot, thinqspace]{SIunits}

\renewcommand{\Re}[1]{\mathrm{Re}\left(#1\right)}
\renewcommand{\Im}[1]{\mathrm{Im}\left(#1\right)}

\usepackage[numbers, sort&compress]{natbib}
\begin{document}	
\title{Sensitive absorption imaging of single atoms in front of a mirror}

\author{Atreju Tauschinsky and Robert J.C. Spreeuw$^*$}
\address{Van der Waals-Zeeman Institute, Institute of Physics, University of Amsterdam, \\
PO Box 94485, 1098 XH Amsterdam, The Netherlands}
\email{$^*$R.J.C.Spreeuw@uva.nl}

\date{\today}

\begin{abstract}
In this paper we show that the sensitivity of absorption imaging of ultracold atoms can be significantly improved by imaging in a standing-wave configuration. We present simulations of single-atom absorption imaging both for a  travelling-wave and a standing-wave imaging setup, based on a scattering approach to calculate the optical density of a single atom. We find that the optical density of a single atom is determined only by the numerical aperture of the imaging system. We determine optimum imaging parameters, taking all relevant sources of noise into account. For reflective imaging we find an improvement of 1.7 in the maximum signal-to-noise ratio can be achieved. This is particularly useful for imaging in the vicinity of an atom chip, where a reflective surface is naturally present.
\end{abstract}


\section{Introduction}
The detection of single atoms with high fidelity is an important requirement for many modern atomic physics experiments, and in particular for quantum information science. Single particle imaging has been achieved using fluorescence imaging of ions~\cite{Neuhauser1980} and neutral atoms~\cite{Schlosser2001, Nelson2007, Wilzbach2009, Bucker2009, Heine2010}, but is difficult in atom-chip experiments due to a scattering background from the chip~\cite{Wilzbach2009}. A different approach is to enhance the atom-photon interaction using resonant cavities~\cite{Goldwin2011, Bochmann2010, Gehr2010}. However, free-space absorption imaging of single particles~\cite{Wineland1987, Streed2012, Tey2009} is still a daunting task.

In this paper we propose absorption imaging in a standing wave as a simple and effective way to improve the signal-to-noise ratio (SNR) of atom detection. This is particularly useful in the context of atom-chip experiments~\cite{Smith2011} where a standing wave naturally forms if the imaging is done perpendicular to the chip surface. We will take the setup currently in use in our group as a reference~\cite{Leung2011, Gerritsma2007, Whitlock2009} for realistic experimental parameters.

In the first section we will calculate the absorption signal of a single atom, both for traditional one-pass imaging and for reflection imaging as employed in our experiment. 
In the second half of the paper we present numerical simulations of the expected absorption images and signal-to-noise ratio (SNR) under realistic conditions, taking into account all common sources of noise, for in-trap atoms as well as for untrapped atoms. To do so we use a hybrid approach, first calculating the movement of the atom in the imaging beam based on scattering individual photons, and subsequently investigating the absorption signal due to this atom, now taking the imaging beam as a classical wave. We show that using reflection imaging drastically increases the expected SNR and thus the single-atom detection accuracy.

\section{Single-Atom Absorption Imaging}
Our starting point is the absorption signal of a single atom at rest, illuminated by a resonant laser beam with homogeneous intensity distribution (see Fig. \ref{fig:setup}). As is shown in the appendix the intensity in the imaging plane is then given by
\begin{equation}
	\frac{I_\mathrm{abs}(\rho)}{I_\mathrm{in}(\rho)}=\frac{\sigma}{1+s}\,\left[\Re{p(\rho)}-a\,\chi\,|p(\rho)|^2\right]
\end{equation}
where $I_\mathrm{abs}(\rho)$ is the absorbed intensity, i.e. the missing intensity in the image plane with image-plane coordinates $\rho$, $I_\mathrm{in}$ is the incoming intensity, $\sigma$ is the absorption cross section of the atom, $s$ the saturation parameter in the object plane, $p(\rho)$ the point-spread function of the imaging system (see below), $a=\left[\int{|p(\rho)|^2\,d^2\rho}\right]^{-1}$ the effective area over which the signal is distributed in the detection plane and $\chi$ is the fraction of the scattered light collected by the imaging system.

In an experiment the above quantities $a$ and $I_\mathrm{abs}(0)$ are not directly accessible due to the finite size of any detector and the finite time resolution. Instead one would observe the quantities 
\begin{equation}
\begin{split}
N_\mathrm{det}(x) &= \int_Ad\rho\int_0^\tau dt\left(I_\mathrm{in}-I_\mathrm{abs}\right)\\
N_\mathrm{ref}(x) &= \int_Ad\rho\int_0^\tau dt I_\mathrm{in}
\label{eq:Nphot}
\end{split}
\end{equation}
where $x$ is an index over camera pixels,  $A$ is the size of a pixel and $\tau$ is the exposure time, i.e. the duration of the imaging pulse. Both $\sigma$ and $s$ now depend on time, as the atom acquires a finite velocity during the imaging pulse, leading to Doppler shifts. In addition the local saturation parameter in the object plane varies in the case of standing wave imaging, as will be discussed below. We can then define the apparent column density of atoms per pixel in the image plane as
\begin{equation}
n(x) = \frac{1+s_0}{\sigma_0}\frac{N_\mathrm{det}(x)}{N_\mathrm{ref}(x)}
\label{eq:atomdensity}
\end{equation}
where $s_0$ is the saturation parameter for the intensity of the incoming beam and $\sigma_0$ is the absorption cross section of an atom at rest for the incoming intensity. The apparent atom number per pixel would be given by $N_\mathrm{app}(x) = A\,n(x)$. 

As the atoms will move out of the focal plane during the imaging pulse, simulating this signal will require a point-spread function that is a function not only of the image plane coordinates, but also takes defocussing into account. This PSF can be numerically determined as
\begin{equation}
p\left(\rho, f; t\right) = \frac{2\pi}{\rho_0^2}\int_{0}^{1}r\,\mathrm{exp}(i f r^2)\times J_0\left(\nicefrac{2\pi\rho\,r}{\rho_0}\right)dr
\label{eq:psf}
\end{equation}
for any given defocus $f=\frac{2\pi}{\lambda}Z(1-\sqrt{1-\mathrm{NA}^2})$ where Z is a real-space coordinate in the imaging direction~\cite{Janssen2002}, NA is the numerical aperture of the imaging system and both $\rho$ and $Z$ might depend on time. $\rho_0=\lambda/\mathrm{NA}$ is proportional to the resolution of the imaging system, determined by the wavelength and the numerical aperture.  Aberrations in the imaging system can be taken into account in a similar manner. We assume these to be negligible near the optical axis. In the simulations the numerical point-spread function given above is used to calculate the intensity in the image plane.

\subsection{Analytical Estimates}
\label{sec:analytics}
For a better intuition of the above results we consider some simple analytic estimates. Assuming a stationary atom in perfect focus of a two-lens imaging system with unit magnification ($f$-$2f$-$f$) we can use a simple point-spread function of the form $p(\rho) = 1/(\rho\rho_0)J_1(2\pi\rho/\rho_0)$ where $p(0)=\pi/\rho_0^2=a^{-1}$. Using $\chi=(\nicefrac{3}{8})\mathrm{NA}^2$ and assuming a 2-level system where $\sigma_0=3\lambda^2/2\pi$ the maximum signal at the center of the image simplifies to
\begin{align}
\frac{I_\mathrm{abs}(0)}{I_\mathrm{in}(0)} &= \frac{1}{1+s_0}\left(\frac{3}{2}\mathrm{NA}^2-\frac{9}{16}\mathrm{NA}^4\right)
\label{eq:intensity_estimate}
\end{align}
which is thus determined purely by the numerical aperture of the imaging system and the saturation parameter $s_0$. For a numerical aperture $\mathrm{NA}=0.4$ as is the case in our experiment we then find an absorption in the centre of the image of $I_\mathrm{abs}(0)/I_\mathrm{in}(0)=0.23$ in the low saturation limit $s_0\ll1$ for a single atom.

We can further consider the total apparent number of atoms one would extract from this signal as the peak signal times the area over which the signal is distributed,
\begin{equation}
N_\mathrm{app} = a\frac{1+s_0}{\sigma_0}\frac{I_\mathrm{abs}(0)}{I_\mathrm{in}(0)} = 1-\frac{3}{8}\mathrm{NA}^2
\label{eq:napp}
\end{equation}
This shows that the expected number of atoms is reduced by the second, NA-dependent term. This term is due to light which is scattered by the atom, but scattered into the solid angle of the lens, thus reducing the apparent amount of light that is absorbed. For a numerical aperture of 0.4 we therefore expect to find an apparent atom number $N_\mathrm{app}$ of 0.94 rather than one.

The commonly used quantity \emph{optical density} (OD) can thus be defined for a single particle by the point-spread function of the imaging system, i.e. $\mathrm{OD} = -\ln\left(1-\nicefrac{I_\mathrm{abs}}{I_\mathrm{in}}\right)$. At the center of the image (for vanishing pixel size) we find for the peak optical density a value of $\mathrm{OD}(0) \approx \nicefrac{3}{2}\mathrm{NA}^2+\nicefrac{9}{16}\mathrm{NA}^4$ in the low-saturation limit (where the sign change is due to a series expansion of the logarithm). For our numerical aperture this equals a peak optical density of 0.26.
It is worth noting here that the atom number extracted in the usual way from optical density, in this case
\begin{equation}
N_\mathrm{OD}=-a\frac{1+s_0}{\sigma_0}\,\ln\left(1-\frac{I_\mathrm{abs}\left(0\right)}{I_\mathrm{in}\left(0\right)}\right)
\label{eq:nod}
\end{equation} would yield 1.07 atoms rather than 0.94. The atom number is slightly overestimated, as Lambert-Beers law is valid only for a continuously absorbing medium, not for a single absorber. In the following we always use the spatially dependent equivalent of \eqref{eq:napp} rather than \eqref{eq:nod} for the atom number.

\subsection{Reflection Imaging}

\begin{figure}[htb]
\centering
\includegraphics[width=7cm]{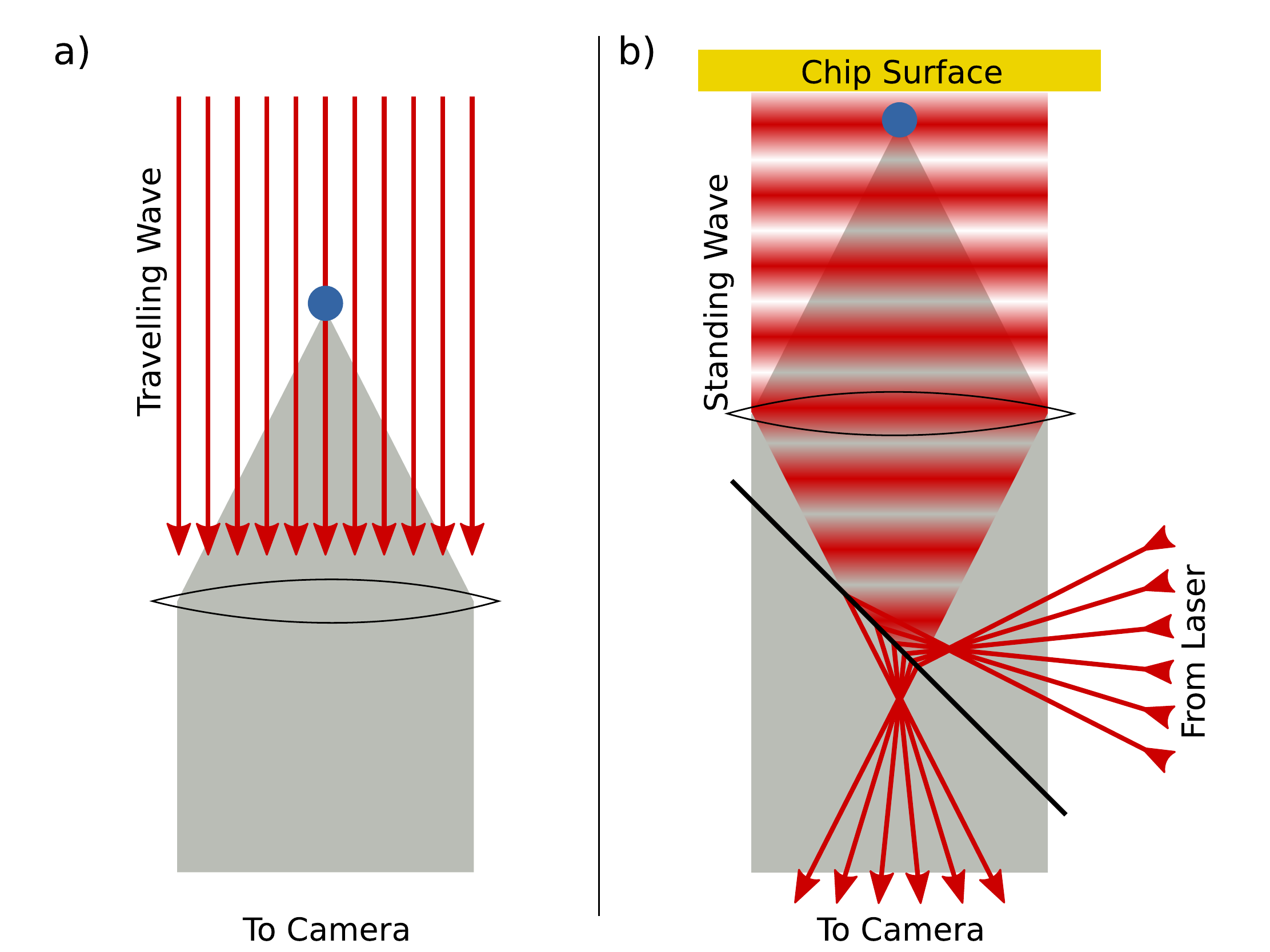}%
\caption{Sketch of the imaging system considered for this paper. On the left the situation of a travelling wave without reflection is depicted. On the right the situation of reflective imaging (with the atom chip on top) is shown.}%
\label{fig:setup}%
\end{figure}

The above discussion is for an atom imaged by a travelling wave,  as  depicted in Fig.~\ref{fig:setup}(a). In reflection imaging, such as in Fig.~\ref{fig:setup}(b) the situation is slightly different. In this case the probe light forms a standing wave at the position of the atoms. For Rb atoms in our magnetic lattice~\cite{Leung2011, Gerritsma2007, Whitlock2009} we expect a ground-state size of the atomic wavefunction of about \unit{40}{\nano\meter}, about a factor of $10$ smaller than the period of the standing wave. We therefore approximate the atom as a point-particle which can initially be positioned in an anti-node of the standing wave. In our atom chip experiment the trap position can be adjusted by means of an externally controlled magnetic field to achieve this.

The intensity of the probe beam then varies sinusoidally along the probe direction and the maximum intensity in the anti-nodes is a factor four greater than the intensity of the incoming travelling wave. The coherent scattering amplitude $\mathcal{A}$ scales with the the local field amplitude (at low saturation). Thus the scattered wave is two times as strong as for a travelling wave, leading to an accordingly higher observed signal. More precisely, the saturation parameter will also change. 

It should be noted here that while in the limit of low saturation and short exposure times the observed signal will indeed be two times larger than for a travelling wave, for realistic imaging with finite exposure time the atom starts to probe the spatial varation of the light field, leading to an observed signal which depends on the exact imaging parameters. The expected signal can then only be predicted by simulations such as described below.

A further important difference to travelling wave imaging is that the atom is not pushed out of the focus of the imaging system by the probe beam, but performs a random walk in all dimensions. This point will also be further discussed in the simulations below.

Another factor of two could in principle be gained because the  scattered wave is emitted towards and reflected by the mirror. This again doubles the amplitude of the scattered wave in the detection plane. A practical consideration here is that for single-atom imaging one will typically use high-NA imaging, with a Rayleigh length $z_R=\lambda/\left(\pi\mathrm{NA}^2\right) \lesssim \unit{10}{\micro\meter}$. For the typical situation in our experiment the atom is at about \unit{10}{\micro\meter} from the surface. This implies that the atom and its mirror image are not simultaneously in focus in the image plane. In practice this reflected wave hardly contributes to the optical density for our experimental parameters. If the imaging is done extremely close to a reflecting surface, or if the Rayleigh length of the imaging system is significantly larger this can however be a significant effect.

\section{Simulations}
\label{sec:simulations}
After having calculated the absorption signal of a single atom at rest in the previous section, we would now like to determine the accuracy with which we can detect a single atom in a more realistic setup. To do so we have to take two important effects into account: recoil blurring and imaging noise. The first describes the process where the atom starts to move due to scattering photons, leading to a blurring of the image. The second describes the uncertainty with which one can determine the local intensity of the imaging beam, both for fundamental reasons, i.e. shot-noise, and technical reasons such as read-out noise of the camera. These two cases impose counteracting constraints: recoil blurring is minimal for short exposure times and weak probe intensities, shot-noise is reduced for large photon counts, i.e. long exposure times and high probe intensities. In addition, at high intensities the signal saturates. Therefore there exists an optimum signal-to-noise ratio (SNR) for a certain exposure time and probe intensity. We use a hybrid approach in treating these effects: recoil blurring is calculated based on scattering individual photons off the atom (see below), while the absorption signal is calculated based on a classical-wave approach as described above.

In our simulations we distinguish four cases:
\begin{description}
	\item[a)] free atom in travelling wave
	\item[b)] trapped atom in travelling wave
	\item[c)] free atom in standing wave
	\item[d)] trapped atom in standing wave
\end{description}

where the cases a) and b) are for traditional one-pass imaging, and the cases c) and d) are for our reflective atom-chip setup (see Fig. \ref{fig:setup}). While case a) and c) assume a freely moving atom, case b) and d) assume the atom to be held in a magnetic trap. For this trap we assume the potential calculated from the magnetic pattern of microtraps present on our atom chip~\cite{Leung2011} which is approximately equivalent to trap frequencies of $(\omega_x, \omega_y, \omega_z) = 2\pi\times$\unit{(38.1, 36.5, 14.0)}{\kilo\hertz} and a trap depth of \unit{8.8}{G}. In the simulations the calculated potential is used without harmonic approximations. The atom is treated as free as soon as it acquires sufficient energy to escape the trap. We assume optical pumping to non-trapped states is negligible for an appropriate choice of polarisation for the imaging beam.

\begin{table}
\centering
\begin{tabular}{cccccccc}
case & $\tau$ ($\mu$s) & $I/I_\mathrm{sat}$  & SNR$_\mathrm{px}$ & SNR$_\mathrm{CRB}$ & N$_\mathrm{app}$ & $p_{1, 1}$ & $p_{1, 0}$\\\hline
	a) & 12.4            & 0.59 							 & 1.55              & 1.57								& 0.94 & 57\% & 21\%\\
	b) & 12.0            & 0.65 							 & 1.57              & 1.59								& 0.95 & 57\% & 20\%\\
	c) & 17.0            & 0.56 							 & 1.73              & 1.76								& 0.74 & 62\% & 19\%\\
	d) & 42.5            & 0.56 							 & 2.63              & 2.68								& 0.83 & 82\% & 9.5\%\\
\end{tabular}
\caption{Optimum exposure parameters, and resulting SNR for the four cases described in the text. SNR$_\mathrm{px}$ is the SNR obtainable from evaluating a single pixel, while SNR$_\mathrm{CRB}$ is the SNR for an estimator achieving the Cramér-Rao bound (without taking fringe-removal into account). The normalization factor N$_\mathrm{app}$ for case a) and b) is as expected from the analytic treatment of section \ref{sec:analytics}. The normalization factor in cases c) and d) can only be obtained from the simulations. Finally we give the probability of true positive measurements $p_{1, 1}$ and false positive measurements due to the presence of zero atoms $p_{1, 0}$ (see \ref{sec:noise}).}
\label{tab:parms}
\end{table}

\subsection{Recoil Blurring}
As the atom is illuminated by the imaging beam it scatters photons at a rate determined by the local intensity at the position of the atom as well as any detuning due to e.g. Doppler shifts as the atom acquires momentum with each of these scattering events. In the case of a travelling wave the photon scattering drives the atom strongly along the direction of the imaging beam, as the absorption of a photon always happens in this direction, performing a biased random walk. This quickly leads to the atom moving out of the focal plane of the imaging lens. The atom further performs an unbiased random walk in momentum space in the plane perpendicular to the imaging direction due to the spontaneous emission of a scattered photon, leading to a blurring of the image. In the case of a standing wave the atom can absorb a photon from both directions along the imaging axis, giving rise to an unbiased random walk in the imaging direction as well as the plane perpendicular to it. The atom therefore leaves the focus of the lens much more slowly than in the case of a travelling wave.

We simulate the movement of the atoms by calculating stochastic trajectories where random scattering events occur at the local scattering rate, taking the effect of the trap into account as a classical potential. In the case of an untrapped travelling wave we recover the expected behaviour, i.e. the root-mean-square (RMS) position along the imaging beam growing as $t^2$ and the RMS position in a plane perpendicular to the imaging direction growing as $t^{3/2}$, as is expected for uniform acceleration and a random walk in momentum space respectively. The effect of the trap is almost negligible in the case of a travelling wave, as the force of the light field is much greater than the restoring force of the trap potential. In the case of reflective imaging the behaviour proportional to $t^{3/2}$ is also expected in the probe direction. This is however modified by the periodic intensity of the standing wave. Since the net force on the atoms is much lower than in the case of a travelling wave the influence of the trapping field is also more important, particularly at low intensities. 

These calculated RMS trajectories are then used in the integrations of Eq. \eqref{eq:Nphot} to determine the signal after a given exposure time for a certain probe intensity.

\subsection{Detection Noise}
\label{sec:noise}
To determine the feasibility of absorption imaging for single atoms one needs to consider various sources of noise in addition to the atomic trajectories to determine the signal-to-noise ratio. In the following we assume near shot-noise limited imaging, with further contributions from camera read-out noise and dark counts, as well as a finite quantum efficiency of the imaging system. Equations \eqref{eq:Nphot} determine the number of photons per pixel in the signal and light fields, and can be trivially modified to take a finite quantum-efficiency of the imaging system into account. With these values we can calculate the expected signal $n(x)$ as atom density per pixel (Eq. \ref{eq:atomdensity}) for any combination of exposure time and intensity. The variance of this signal is given by

\begin{equation}
\sigma_n(x)^2 = \left(\frac{1+s_0}{\sigma_0}\right)^2\left(\frac{\sigma_\mathrm{det}^2}{N_\mathrm{ref}^2} + \frac{N_\mathrm{det}^2\sigma_\mathrm{ref}^2}{N_\mathrm{ref}^4}\right)
\end{equation}

where $N_\mathrm{det}$ and $N_\mathrm{ref}$ are the number of electrons per pixel in the signal and reference images respectively, also integrated over pixel size and exposure time. $\sigma_\mathrm{det}^2$ and $\sigma_\mathrm{ref}^2$ are the variances of these quantities, mostly determined by photon shot noise ($\propto N$) and camera readout noise.

The signal to noise ratio per pixel is then given by $\nicefrac{n(x)}{\sigma_n(x)}$; one can further improve the signal to noise ratio by not only evaluating the central pixel, but using an optimal estimator achieving the Cramér-Rao bound, such as 
\begin{equation}
N = \frac{\sum{n(x) q(x)}}{\sum q(x)^2}
\label{eq:estimator}
\end{equation}
where $q(x)$ is the spatial mode function of the signal~\cite{Ockeloen2010}. In our results we list the SNR both for evaluating only the central pixel and for an estimator achieving the Cramér-Rao bound.

We define the accuracy of the measurement as the probability of finding true positives $p_{1, 1}$, i.e. finding one atom if there is one atom present. Here we assume all measurement outcomes for which $0.5 < \mathrm{N} < 1.5$ to indicate the presence of exactly one atom. We have determined this value by simulations of $10^6$ individual absorption images for each case. The results are in excellent agreement with those expected for a normal distribution, for which the accuracy is given by $F=\mbox{Erf}(\nicefrac{z}{\sqrt{2}})$ with $z=\nicefrac{N}{2\sigma}$.

We can also determine the probability of finding false positives in the absence of any atoms $p_{1, 0}$, i.e. finding one atom if zero atoms are present. False positives due to the presence of more than one atom can not easily be determined, as an accurate calculating of the absorption signal of two atoms is non-trivial. We expect these to be of similar magnitude as those due to the absence of any atoms.

\subsection{Simulation Results}

\begin{figure}[ht]
	\centering
	\includegraphics[width=7cm]{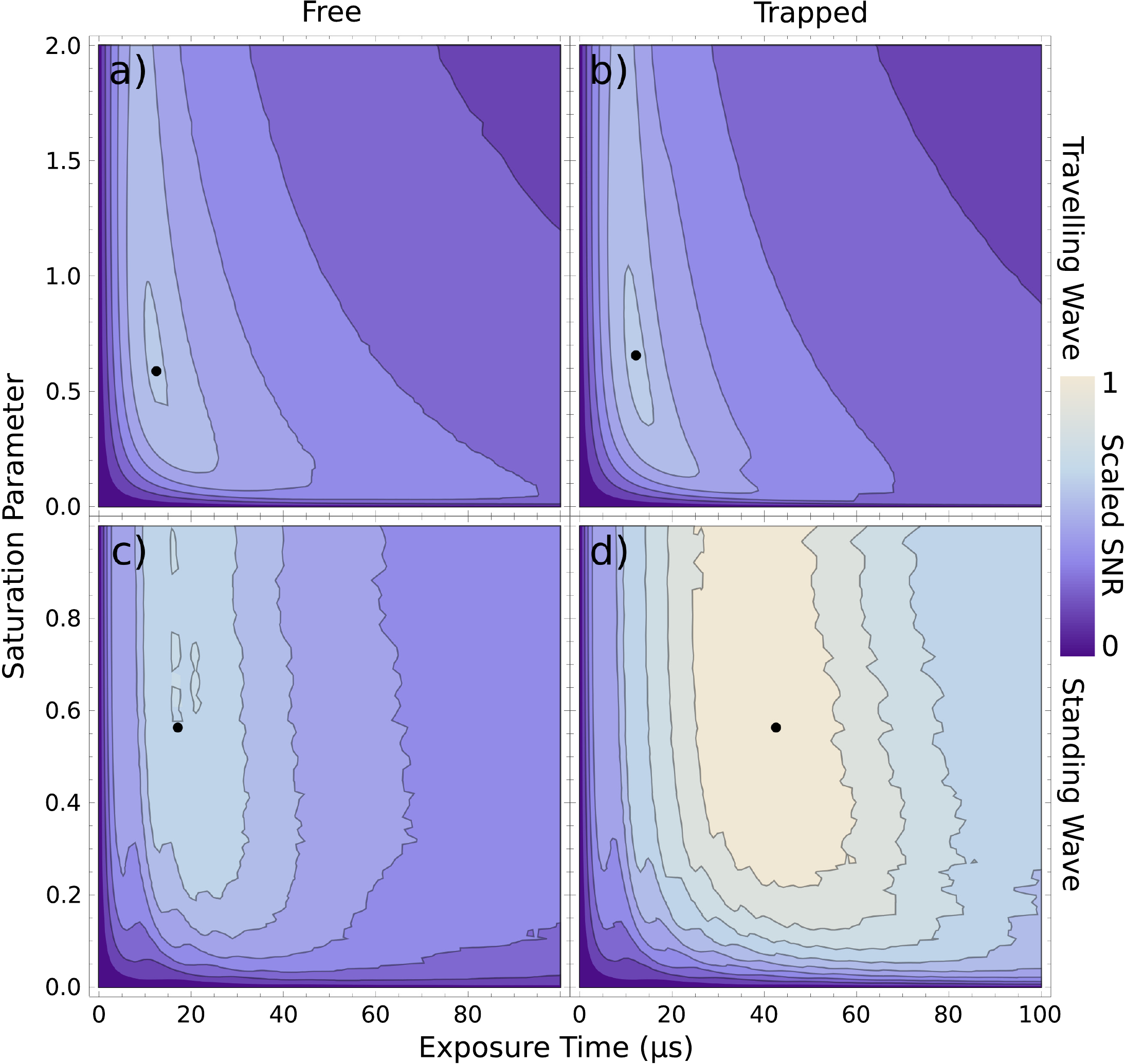}
	\caption{Simulated SNR as function of intensity of the incoming beam (in units of saturation intensity) and exposure time. The four subplots correspond to the four cases described in the text as indicated near the top and right axes, the color scale is normalized to the maximum SNR of case d). The black dot marks the position of the optimum.}
	\label{fig:SNR}
\end{figure}

\begin{figure}[ht]
	\centering
	\includegraphics[width=7cm]{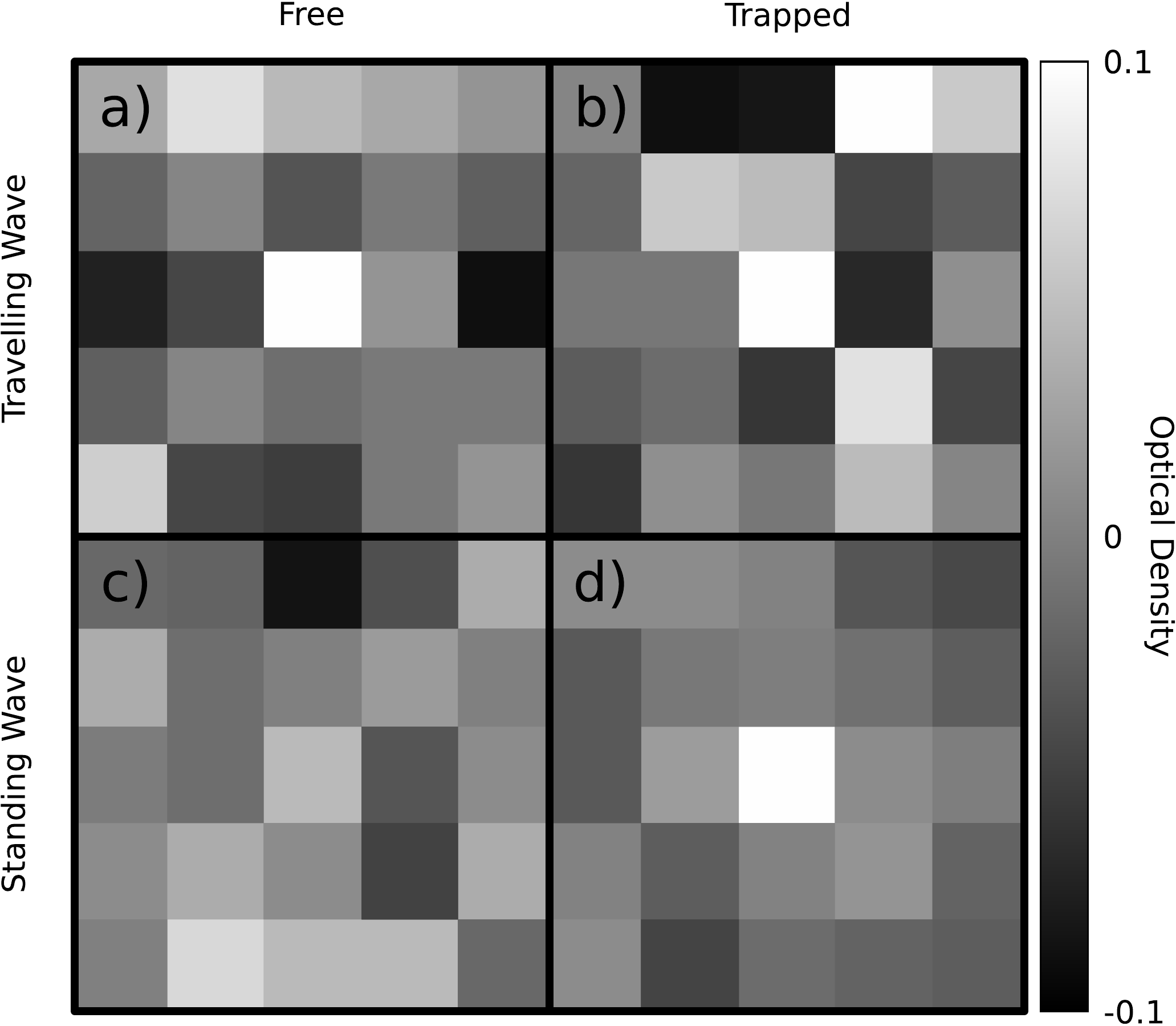}
	\caption{Simulated single-shot absorption images for optimum exposure parameters, including fringe removal. The four subplots correspond to the four cases described in section \ref{sec:simulations}. Bright colors correspond to high optical density. Each subplot shows $5\times5$ pixels, with the atom initially located at the center. At larger distances from the initial position the PSF becomes negligibly small. Using Eq. \ref{eq:estimator} we extract N=(1.32, 1.39, 0.56, 1.26) atoms for subplots a)-d) respectively. These atom numbers are normalized by $N_\mathrm{app}$ for each imaging setup.}
	\label{fig:od}
\end{figure}

We simulate the expected absorption image and SNR assuming a numerical aperture of the objective lens of 0.4 (Edmund Optics NT47-727 with custom AR coating), a total magnification of the imaging system of 10, a pixel size of the camera of $\unit{(13 \times 13)}{\micro\meter\squared}$ (Andor iKon-M 934), a quantum efficiency of 0.9 and a readout noise of 13 counts per pixel~\cite{Ockeloen2010}. We consider resonant $\sigma^+$-transitions on the D2-line of Rubidium 87, where $\sigma_0=\unit{2.907\times10^{-9}}{\centi\meter\squared}$ and the saturation intensity $I_\mathrm{sat}=\unit{1.669}{\milli\watt\per\centi\meter\squared}$~\cite{steck2010}. We have confirmed that in the limit of low saturation our simulations yield the expected number of atoms in all four cases.

We find optimum imaging parameters by numerically maximizing the signal-to-noise ratio of the central pixel as a function of exposure time and intensity, i.e. maximizing SNR(0) where 
\begin{equation}
\mathrm{SNR}(x) = \nicefrac{n(x)}{\sigma_n(x)}
\end{equation}
using the precalculated trajectories of an atom in the integrations of Eqs. \eqref{eq:Nphot}.
To avoid confusion in the case of reflective imaging where the intensity is spatially varying, the intensity is here always given in units of saturation intensity for the incoming beam (cf. Fig. \ref{fig:setup}). The local intensity at the initial position of the atoms is therefore higher by a factor of four for the standing wave. 
We also determine the SNR of an estimator achieving the Cramér-Rao bound by weighting the signal by the mode function of the atomic distribution as described above. SNR as a function of exposure time and intensity is depicted in Fig. \ref{fig:SNR}. Table \ref{tab:parms} lists the signal-to noise ratio for optimum imaging parameters. Note that this can be further improved by a factor $\approx \sqrt{2}$ by using the \emph{fringe-removal algorithm}~\cite{Ockeloen2010} which reduces the shot noise in the light field by optimal averaging of reference images even in the absence of any fringes. Fig. \ref{fig:od} finally shows simulated single-shot absorption images of a single atom for the parameters given above, including the use of fringe removal. The number of atoms extracted from these images using the optimum estimator is given in the figure caption.

Depending on the imaging method the effect of recoil blurring on the optical resolution of the imaging system can be significant. For our optimum exposure parameters we find the strongest recoil blurring in case  c), where the RMS position of the atoms in the object plane increases by \unit{0.9}{\micro\meter} during the imaging pulse. For all other cases considered here recoil blurring is smaller by at least a factor of 3. This blurring should be compared to the base resolution of the imaging system of \unit{1.2}{\micro\meter} (Rayleigh criterion). Furthermore the images are discretized by the effective pixel size of \unit{1.3}{\micro\meter} in object space. 

\section{Discussion \& Conclusion}
In this paper we have outlined the theoretical prospects of detecting single rubidium atoms using absorption imaging. We find the optical density of a single atom in single-pass imaging to be ultimately determined only by the numerical aperture of the imaging system with a peak optical density of $OD = \nicefrac{3}{2}\mathrm{NA}^2\left(1+\nicefrac{3}{8}\mathrm{NA}^2\right)$ in the low saturation limit. We have calculated optimum imaging parameters for a number of different cases and could show that the use of a reflective imaging setup can significantly improve the signal-to-noise ratio of absorption imaging. We expect an accuracy of about $82\%$ for the detection of single atoms for the optimum parameters given above in case d), which can be improved to $94\%$ (with a false-positive rate due to zero atoms of $4\%$) using the reduction in shot noise achieved by the fringe removal algorithm~\cite{Whitlock2010}. This compares to a detection accuracy of only $74\%$ (with a false-positive rate due to zero atoms of $15\%$) for single-pass imaging with the same imaging system. For measuring statistical averages in a regular array of traps further improvements can be made using correlation analysis~\cite{Whitlock2010}.

We would like to thank Jasper Reinders for help with the simulations and N.J. van Druten for helpful discussions of the manuscript.
This work is part of the research programme of the Foundation for Fundamental Research on Matter (FOM), which is part of the Netherlands Organisation for Scientific Research (NWO).

\appendix
\section{Absorption Signal of a Single Atom}

We start from the  simple configuration of a single atom at rest, placed in the centre of the minimum waist of a travelling-wave laser beam, see Fig.~\ref{fig:setup}(a). 
The laser is taken as monochromatic,  $\mathbf{E}_\mathrm{in}(x_o,y_o,z,t)=\frac{1}{2}\mathcal{E}_\mathrm{in}^\mathrm{(o)}(x_o,y_o)\mathbf{u}\,e^{i(kz-\omega t)}+c.c.$, where $c.c.$ denotes complex conjugate, $\mathbf{u}$ is a complex polarization unit vector, $\mathcal{E}_\mathrm{in}^\mathrm{(o)}(x_o,y_o)$ is the amplitude in the object plane with coordinates $x_o$, $y_o$. We assume that the laser is collimated in the object plane, so we can choose $\mathcal{E}_\mathrm{in}^\mathrm{(o)}(x_o,y_o)$ to be real. The incident intensity is $I_\mathrm{in}^\mathrm{(o)}(x_o,y_o)=[\mathcal{E}_\mathrm{in}^\mathrm{(o)}(x_o,y_o)]^2/2Z_0$, with $Z_0=(\epsilon_0 c)^{-1}=\sqrt{\mu_0/\epsilon_0}$.

We describe the total light field as a sum of this incident plane wave plus a scattered wave. Assuming they have the same polarisation we ignore from now on the polarization vector. The total light field is imaged onto a detector plane where we measure the total intensity $I_\mathrm{det}(\rho=(x,y))$. 
The scattered wave has both a coherent and an incoherent contribution, which we add separately as follows, 
\begin{align}
I_\mathrm{det}(\rho)
&=\frac{1}{2Z_0}|\mathcal{E}_\mathrm{in}(\rho)+\mathcal{E}_{\mathrm{sc}}^{(\mathrm{coh})}(\rho)|^2+I_{\mathrm{sc}}^{(\mathrm{incoh})}(\rho)\\
&=I_\mathrm{in}(\rho)+I_{\mathrm{sc}}(\rho)+I_{\mathrm{if}}(\rho)
\label{eq:irho}
\end{align}
The total scattered intensity $I_{\mathrm{sc}}(\rho)$ is the sum of a coherent and an incoherent contribution, $I_{\mathrm{sc}}^{(\mathrm{incoh})}(\rho)=sI_{\mathrm{sc}}^{(\mathrm{coh})}(\rho)$ where $s$ is the saturation parameter~\cite{Cohen-Tannoudij1992}. The third term is the interference term,
\begin{equation}
	I_{\mathrm{if}}(\rho)=\frac{\mathcal{E}_\mathrm{in}(\rho)}{Z_0}\,\Re{\mathcal{E}_{\mathrm{sc}}^{(\mathrm{coh})}(\rho)}
	\label{eq:Iif}
\end{equation}
where $\mathcal{E}_{\mathrm{sc}}^{(\mathrm{coh})}(\rho)$ is complex-valued. The interference intensity is typically negative at the center of the image, as it reduces the total intensity in the image plane.

The absorbed intensity is given by
\begin{equation}
	I_\mathrm{abs}(\rho) =I_\mathrm{in}(\rho)-I_\mathrm{det}(\rho)=-I_\mathrm{if}(\rho)-I_\mathrm{sc}(\rho)
	\label{eq:iabs}
\end{equation}
The terms on the right can be related to the total scattered power $P_{\mathrm{sc}}$ after integration. 
The imaging system collects a fraction $\chi$ of the scattered power, while a fraction $1-\chi$ is not collected. The latter represents the power that is ``missing'' in the detection plane, by virtue of energy conservation:
\begin{align}
	\chi P_{\mathrm{sc}} &=\int{I_\mathrm{sc}(\rho)\,d^2\rho}\\
	(1-\chi) P_{\mathrm{sc}} &=\int{\left[I_\mathrm{in}\left(\rho\right)-I_\mathrm{det}\left(\rho\right)\right]\,d^2\rho}
\end{align}
where the integration is over the detection plane. These expressions presume that the incident laser beam is entirely captured by the imaging system, so that no interference occurs in the noncollected light. When we add  the scattered fractions  only the interference integral remains, 
\begin{equation}
	P_\mathrm{sc}=-\int{I_\mathrm{if}(\rho)\,d^2\rho}
	\label{eq:Psc}
\end{equation}
The interference integral must clearly be negative, and equal to minus the total scattered power (coherent plus incoherent). Note that it is independent of the collection solid angle. 

Next, we introduce the point spread function to describe how the point-like atom is imaged in the detection plane:
\[
	\mathcal{E}_{\mathrm{sc}}^{(\mathrm{coh})}(\rho)=\mathcal{A}\, p(\rho)
\]
with  complex amplitude $\mathcal{A}$. 
We choose the (complex valued) point spread function (PSF) to be normalized as $\int{p(\rho)\,d^2\rho}=1$ so that $p(\rho)$ has the dimension 1/area and $a=\left[\int{|p(\rho)|^2\,d^2\rho}\right]^{-1}$ can be interpreted as the effective area over which the scattered light is distributed in the detection plane. 

The absorption signal, Eq.~\eqref{eq:iabs}, can now be written as 
\begin{equation}
	I_\mathrm{abs}(\rho)=-\frac{\mathcal{E}_\mathrm{in}(\rho)}{2Z_0}\left[\mathcal{A}\,p(\rho)+c.c.\right]-\frac{1+s}{2Z_0}\,|\mathcal{A}|^2\,|p(\rho)|^2
\end{equation}
In most practical cases we can approximate the probe beam as a nearly plane wave, and $p(\rho)$ as sharply peaked compared to $\mathcal{E}_\mathrm{in}(\rho)$. Taking the latter outside the interference integral, Eq.~\eqref{eq:Psc},  replacing $\mathcal{E}_\mathrm{in}(\rho)\approx\mathcal{E}_\mathrm{in}(0)$, we obtain, using Eq.~\eqref{eq:Iif}, 
\begin{equation}
	\int{I_\mathrm{if}(\rho)\,d^2\rho}=\frac{\mathcal{E}_\mathrm{in}(0)\,\Re{\mathcal{A}}}{Z_0}=-P_\mathrm{sc}
\end{equation}

Using the integrals defined above, we can then express the interference term as 
\begin{equation}
	-I_\mathrm{if}(\rho) \approx P_\mathrm{sc}\,\left[\Re{p(\rho)}-\frac{\Im{\mathcal{A}}}{\Re{\mathcal{A}}}\,\Im{p(\rho)}\right]
\end{equation}
If the atom is in the focal plane of the imaging system we can take $p(\rho)$ to be real-valued so that the  $\Im{p(\rho)}$ term vanishes. The term also vanishes if the probe beam is exactly on resonance with the atomic transition. In this case $\Im{p(\rho)}/\Re{p(\rho)}=\tan(\arg\mathcal{A})=0$. The approximation of a resonant probe beam is valid as all Doppler shifts are less than 2\% of the natural linewidth of the transition for the optimum exposure parameters found in the main part of the paper. In both cases the absorption signal then reduces to 
\begin{equation}
	I_\mathrm{abs}(\rho) \approx P_\mathrm{sc}\left[\Re{p(\rho)}-a\,\chi\,|p(\rho)|^2\right]
\end{equation}
The leading term in the absorption is simply given by the scattered power. The correction expresses the partial ``refilling'' of the absorption dip by scattered light. 
Within the above approximation the absorption signal can also be expressed in terms of the scattering cross section, using $P_\mathrm{sc}=\sigma I_\mathrm{in}(0)/(1+s)$, 
\begin{equation}
	\frac{I_\mathrm{abs}(\rho)}{I_\mathrm{in}(0)}=\frac{\sigma}{1+s}\,\left[\Re{p(\rho)}-a\,\chi\,|p(\rho)|^2\right]
\end{equation}


\begin{thebibliography}{10}
\newcommand{\enquote}[1]{``#1''}

\bibitem{Neuhauser1980}
W.~Neuhauser, M.~Hohenstatt, P.~Toschek, and H.~Dehmelt, \enquote{{Localized
  visible Ba\^{}\{+\} mono-ion oscillator},} Physical Review A \textbf{22},
  1137--1140 (1980).

\bibitem{Schlosser2001}
N.~Schlosser, G.~Reymond, I.~Protsenko, and P.~Grangier,
  \enquote{{Sub-poissonian loading of single atoms in a microscopic dipole
  trap.}} Nature \textbf{411}, 1024--7 (2001).

\bibitem{Nelson2007}
K.~D. Nelson, X.~Li, and D.~S. Weiss, \enquote{{Imaging single atoms in a
  three-dimensional array},} Nature Physics \textbf{3}, 556--560 (2007).

\bibitem{Wilzbach2009}
M.~Wilzbach, D.~Heine, S.~Groth, X.~Liu, T.~Raub, B.~Hessmo, and
  J.~Schmiedmayer, \enquote{{Simple integrated single-atom detector},} Optics
  Letters \textbf{34}, 259 (2009).

\bibitem{Bucker2009}
R.~B\"{u}cker, A.~Perrin, S.~Manz, T.~Betz, C.~Koller, T.~Plisson, J.~Rottmann,
  T.~Schumm, and J.~Schmiedmayer, \enquote{{Single-particle-sensitive imaging
  of freely propagating ultracold atoms},} New Journal of Physics \textbf{11},
  103039 (2009).

\bibitem{Heine2010}
D.~Heine, W.~Rohringer, D.~Fischer, M.~Wilzbach, T.~Raub, S.~Loziczky, X.~Liu,
  S.~Groth, B.~Hessmo, and J.~Schmiedmayer, \enquote{{A single-atom detector
  integrated on an atom chip: fabrication, characterization and application},}
  New Journal of Physics \textbf{12}, 095005 (2010).

\bibitem{Goldwin2011}
J.~Goldwin, M.~Trupke, J.~Kenner, a.~Ratnapala, and E.~a. Hinds, \enquote{{Fast
  cavity-enhanced atom detection with low noise and high fidelity.}} Nature
  communications \textbf{2}, 418 (2011).

\bibitem{Bochmann2010}
J.~Bochmann, M.~M\"{u}cke, C.~Guhl, S.~Ritter, G.~Rempe, and D.~L. Moehring,
  \enquote{{Lossless State Detection of Single Neutral Atoms},} Physical Review
  Letters \textbf{104}, 203601 (2010).

\bibitem{Gehr2010}
R.~Gehr, J.~Volz, G.~Dubois, T.~Steinmetz, Y.~Colombe, B.~L. Lev, R.~Long,
  J.~Est\`{e}ve, and J.~Reichel, \enquote{{Cavity-Based Single Atom Preparation
  and High-Fidelity Hyperfine State Readout},} Physical Review Letters
  \textbf{104}, 203602 (2010).

\bibitem{Wineland1987}
D.~J. Wineland, W.~M. Itano, and J.~C. Bergquist, \enquote{{Absorption
  spectroscopy at the limit: detection of a single atom.}} Optics letters
  \textbf{12}, 389--91 (1987).

\bibitem{Streed2012}
E.~W. Streed, A.~Jechow, B.~G. Norton, and D.~Kielpinski, \enquote{{Absorption
  imaging of a single atom.}} Nature communications \textbf{3}, 933 (2012).

\bibitem{Tey2009}
M.~K. Tey, G.~Maslennikov, T.~{C H Liew}, S.~A. Aljunid, F.~Huber, B.~Chng,
  Z.~Chen, V.~Scarani, and C.~Kurtsiefer, \enquote{{Interfacing light and
  single atoms with a lens},} New Journal of Physics \textbf{11}, 043011
  (2009).

\bibitem{Smith2011}
D.~A. Smith, S.~Aigner, S.~Hofferberth, M.~Gring, M.~Andersson, S.~Wildermuth,
  P.~Kr\"{u}ger, S.~Schneider, T.~Schumm, and J.~Schmiedmayer,
  \enquote{{Absorption imaging of ultracold atoms on atom chips},} Optics
  Express \textbf{19}, 8471 (2011).

\bibitem{Leung2011}
V.~Y.~F. Leung, A.~Tauschinsky, N.~J. Druten, and R.~J.~C. Spreeuw,
  \enquote{{Microtrap arrays on magnetic film atom chips for quantum
  information science},} Quantum Information Processing \textbf{10}, 955--974
  (2011).

\bibitem{Gerritsma2007}
R.~Gerritsma, S.~Whitlock, T.~Fernholz, H.~Schlatter, J.~A. Luigjes, J.-U.
  Thiele, J.~B. Goedkoop, and R.~J.~C. Spreeuw, \enquote{{Lattice of microtraps
  for ultracold atoms based on patterned magnetic films},} Physical Review A
  \textbf{76}, 033408 (2007).

\bibitem{Whitlock2009}
S.~Whitlock, R.~Gerritsma, T.~Fernholz, and R.~J. Spreeuw,
  \enquote{{Two-dimensional array of microtraps with atomic shift register on a
  chip},} New Journal of Physics \textbf{11}, 023021 (2009).

\bibitem{Janssen2002}
A.~J. E.~M. Janssen, \enquote{{Extended Nijboer–Zernike approach for the
  computation of optical point-spread functions},} Journal of the Optical
  Society of America A \textbf{19}, 849 (2002).

\bibitem{Ockeloen2010}
C.~Ockeloen, A.~Tauschinsky, R.~Spreeuw, and S.~Whitlock, \enquote{{Detection
  of small atom numbers through image processing},} Physical Review A
  \textbf{82}, 061606 (2010).

\bibitem{steck2010}
D.~A. Steck, \enquote{{Rubidium 87 D Line Data},} http://steck.us/alkalidata
  \textbf{2} (2010).

\bibitem{Whitlock2010}
S.~Whitlock, C.~F. Ockeloen, and R.~J.~C. Spreeuw, \enquote{{Sub-Poissonian
  Atom-Number Fluctuations by Three-Body Loss in Mesoscopic Ensembles},}
  Physical Review Letters \textbf{104}, 120402 (2010).

\bibitem{Cohen-Tannoudij1992}
C.~Cohen-Tannoudij, J.~Dupont-Roc, and G.~Grynberg, \emph{{Atom—Photon
  Interactions}} (Wiley-VCH Verlag GmbH, Weinheim, Germany, 1992).

\end{thebibliography}
\end{document}